\documentclass[12pt]{article}
\usepackage{amsmath}
\usepackage{amssymb}
\usepackage{graphicx}
\usepackage{float}
\usepackage{indentfirst}
\usepackage{mathrsfs}
\usepackage{dsfont}

\usepackage{setspace}

\usepackage[labelfont=bf,labelsep=period]{caption}
\usepackage[numbers,sort&compress]{natbib}

\newtheorem{theorem}{Theorem}

\newtheorem{corollary}{Corollary}
\newtheorem{lemma}{Lemma}

\title{\vskip -0.7cm The Optimal Uncertainty Relation \\ [0.7cm]}

\author
{Jun-Li Li and Cong-Feng Qiao$^{\ast}$ \\ [0.2cm]
\normalsize{School of Physical Sciences, University of Chinese Academy of Sciences,} \\
\normalsize{YuQuan Road 19A, Beijing 100049, China}\\[2pt]
\normalsize{Center of Materials Science and Optoelectronics Engineering \& CMSOT,}\\
\normalsize{University of Chinese Academy of Sciences, }\\
\normalsize{ YuQuan Road 19A, Beijing 100049, China} \\ [3mm]
\normalsize{$^\ast$ To whom correspondence should be addressed; E-mail: qiaocf@ucas.ac.cn.}\\ [13mm]
}

\date{}
\date{}

\begin{document}
\baselineskip24pt \maketitle
\begin{abstract} \doublespacing
Employing the lattice theory on majorization, we investigate the universal quantum uncertainty relation for any number observables and general measurement. We find: 1. The least bounds of the universal uncertainty relations can only be properly defined in the lattice theory; 2. Contrary to variance and entropy, the metric induced by the majorization lattice implies an intrinsic structure of the quantum uncertainty; 3. The lattice theory correlates the optimization of uncertainty relation with the entanglement transformation under local quantum operation and classical communication. Interestingly, the optimality of the universal uncertainty relation is found can be mimicked by the Lorenz curve, initially introduced in economics to measure the wealth concentration degree of a society.
\end{abstract}

\newpage

\section{Introduction}

The uncertainty relation may be attributed to the original idea of indeterminacy first proposed by Heisenberg in the form of $p_1 q_1 \sim h$, where $h$ is the Planck constant, $p_1$ and $q_1$ represent the precisions in determining the canonical conjugate observables $p$ and $q$ \cite{Heisenberg-o}. In the literature, whereas the most representative uncertainty relation is the Heisenberg-Robertson one \cite{Robertson}:
\begin{equation}
\Delta X^2 \Delta Y^2 \geq \frac{1}{4}|\langle [X,Y] \rangle|^2 \; . \label{Robertson-Uncertainty}
\end{equation}
Here the uncertainty is characterized in terms of variance ($\Delta X^2$ for an observable $X$). Equation (\ref{Robertson-Uncertainty}) asserts a fundamental limit to the uncertainties of incompatible observables expressed in form of commutator. Besides the product-form, there also exist the sum-form uncertainty relations which will be nontrivial when one of the variances becomes zero, we refer to \cite{Sum-angular, Sum-Qian, Sum-Schwonnek, Sum-Giorda, Sum-Szy} for recent developments along this line.

The essence of different forms of the uncertainty relations lies in the lower bound, whose optimization is generally a challenging task. A lasting criticism on variance based uncertainty relation is about its lower bound state dependence \cite{Entropy-0}. In order to be state independent \cite{Sum-Qian, Reformulating-Li}, the variance based uncertainty relations have to involve complex variance functions \cite{TUR}. On the other hand, the entropic uncertainty relation was proposed with state independent lower bound \cite{Entropy-UN-MU}, in the form
\begin{equation}
H(X)+H(Y) \geq \log_2\frac{1}{c} \; ,
\end{equation}
where $H(X)$ denotes the Shannon entropy of outcome probability distribution while $X$ is measured; $c :=\max_{i,j} |\langle x_i|y_j\rangle|^2$ quantifies the complementarity of observables with $|x_i\rangle$ and $|y_j\rangle$ being the eigenvectors of $X$ and $Y$. Studies indicate that these two different forms of uncertainty relations are in fact mutually convertible \cite{Equiv-V-E}.

One main subject in the study of entropic uncertainty relation is about the lower bound optimization, which turns out to be difficult for general observables in high dimensional system \cite{Entropy-new}. The majorization uncertainty relation has been called universal \cite{Majorization-1} and been exploited to refine the entropic uncertainty relation \cite{Majorization-2}, of which the direct sum form usually has a better lower bound than the direct product ones \cite{Example-RPZ}, and both of them remain to be further optimized \cite{Fei-sci, Fei-improved, Maj-mix-sum}. The majorization relation is a partial order on probability distribution vectors with descending order components, and has been shown to form a lattice \cite{Maj-Latt-1}. The majorization lattice has well defined upper and lower bounds, and a recent development appears in its application to econometrics  \cite{Maj-Latt-2, Maj-Latt-3}. Notice of these, naturally, one is tempted to think of formulating the uncertainty relation from the lattice theory, in order to get a properly defined and optimized uncertainty relation.

In this work, by introducing the lattice theory of quantum uncertainty we derive the optimal universal uncertainty relation in the form of direct-sum majorization relation, which is applicable to multiple observables and general positive operator-valued measurements (POVM). The majorization lattice \cite{Maj-Latt-3} leads us to treat the distribution vectors as ``relative measures'' of quantum uncertainty which are not always comparable and explains why it is difficult to optimize the uncertainty relations involving either variance or entropy. Here the incomparability of the measures means that two distribution vectors may have no particular order in the majorization relation. We illustrate the optimality of the universal uncertainty relation by Lorenz curve that was originally introduced in describing the wealth concentration \cite{Lorenz-curve}. The Lorenz curve is sensitive to the incomparability of quantum uncertainties while the variance and entropy are not, and hence the direct experimental test of the universal uncertainty relations by measuring the Lorenz curves turns out to be feasible.

\section{The optimal universal uncertainty relation}

\subsection{Quantum measurement and the majorization lattice}

In quantum mechanics (QM), physical observables are represented by Hermitian operators. In the $N$-level system, an observable $X$ appears in the form of a $N$-dimensional Hermitian matrix, whose spectrum decomposition goes as
\begin{align}
X = \sum_{i=1}^N x_{i}|x_{i} \rangle \langle x_{i}| \; . \label{Observable-x}
\end{align}
Here, $|x_i\rangle$ is the eigenvector that $X|x_{i}\rangle=x_{i}|x_{i}\rangle$. The quantum state $\rho$ of a system is also a Hermitian matrix with nonnegative eigenvalues $\lambda_i$, which may be expressed as a vector $\vec{\lambda}_{\rho} = (\lambda_1, \ldots, \lambda_N)^{\mathrm{T}}$, where the superscript T denotes the transpose of matrix. Moreover, the measurement postulate of QM tells that when measuring $X$ over a quantum state $\rho$ one can only get its eigenvalue $x_{i}$ with a probability of $p_{i} = \langle x_{i}|\rho| x_{i}\rangle$. Similarly, we can express the probability distribution in form of vector, $\vec{p} = (p_1,\cdots, p_N)^{\mathrm{T}}$.

We define a set of Hermitian operators
\begin{equation}
\mathcal{S}^{(x)}_n = \left\{X_{n} | X_{n} = \sum_{i \in \,\mathcal{I} } \,|x_{i}\rangle \langle x_{i}|, \ \mathcal{I}  \subseteq \{1, \ldots, N\}\, \mathrm{and}\, |\mathcal{I}| = n \right\}\ , \label{Xn-set}
\end{equation}
where $|\cdot |$ means the cardinality of the set $\mathcal{I}$. For given $n$, $|S^{(x)}_n|$ equals to $C({N},{n}) = \frac{N!}{n!(N-n)!}$, that means the operators in $S^{(x)}_n$ are composed of various $n$ distinct projection operators $|x_{i}\rangle\langle x_{i}|$ from the complete set, and evidently $\mathcal{S}_0^{(x)} = \{0\}$. The partial sum of the probability distribution $\vec{p}$ now may be expressed as
\begin{equation}
\sum_{i \in \, \mathcal{I}} p_{i} = \mathrm{Tr}(\sum_{i\in\, \mathcal{I}}|x_i\rangle \langle x_i|\rho) = \mathrm{Tr}[X_n(\mathcal{I}) \rho] \; . \label{partial-sums}
\end{equation}
Here $X_{n}(\mathcal{I})$ denotes the matrix $X_n\in \mathcal{S}_{n}^{(x)}$ specified by the set $\mathcal{I}$. For the POVM, the projection operators $|x_{i}\rangle \langle x_{i}|$ are replaced by positive semidefinite operators $M_{i}$ satisfying the normalization condition $\sum_{i} M_{i}^{\dag} M_{i} = \mathds{1}$ \cite{Quant-Infor-BOOK}, and the probability of outcome $i$ is given by $p_i= \mathrm{Tr}[M_i^{\dag}M_i\rho]$. Hence equation (\ref{partial-sums}) applies to POVM as well.

The majorization relation between two tuples of real numbers, $\vec{a} \prec \vec{b}$ say for instance is defined as \cite{Majorization-Book}:
\begin{equation}
\sum_{i=1}^k {a}^{\downarrow}_i \leq \sum_{j=1}^k {b}^{\downarrow}_j \; , \; k\in \{1,\ldots, N\} \; ,
\end{equation}
where the superscript $\downarrow$ means that the components of vectors $\vec{a}$ and $\vec{b}$ are arrayed in descending order, and the equality holds when $k=N$. For the set
\begin{equation}
\mathcal{P}^{N} = \left\{\vec{p} = (p_1,\cdots, p_N)^{\mathrm{T}}|\, p_i\in [0,1], \sum_{i=1}^Np_i=\mathrm{const.},p_i \geq p_{i+1} \right\} \; ,
\end{equation}
the following Lemma exist \cite{Maj-Latt-1}.
\begin{lemma}
For all $\vec{a},\vec{b} \in \mathcal{P}^{N}$, there exists a unique least upper bound $\vec{u} = \vec{a} \vee \vec{b} \in \mathcal{P}^{N}$ such that the followings are satisfied:\\
1. $\vec{a} \prec \vec{u}$ and $\vec{b} \prec \vec{u}$;\\
2. For arbitrary $\vec{x} \in \mathcal{P}^{N}$, if $\vec{a} \prec \vec{x}$ and $\vec{b} \prec \vec{x}$, then $\vec{u} \prec \vec{x}$. \label{Lemma-1}
\end{lemma}
There also exists a unique greatest lower bound defined as $\vec{a} \wedge \vec{b} \in \mathcal{P}^{N}$, and hence $\mathcal{P}^N$ together with the majorization relation form a lattice. Practical methods for constructing least upper bound $\vec{a}\vee \vec{b}$ and greatest lower bound $\vec{a}\wedge \vec{b}$ are given in Ref. \cite{Maj-Latt-1}.

\subsection{The optimal universal uncertainty relation}

The probability distribution of observable measurement outcomes may be expressed as a high dimensional vector in the form of direct sum. Hence for observables $X$, $Y$, and $Z$, the corresponding vector turns out to be a $3N$-dimensional vector $\vec{\chi} = \vec{p}\,\oplus \vec{q} \,\oplus \vec{r}$, with $p_{i} =\langle x_i| \rho |x_i\rangle$, $q_j=  \langle y_j|\rho |y_j\rangle$, $r_k=\langle z_k| \rho |z_k\rangle$, and one may notice  $\vec{\chi}^{\,\downarrow} \in \mathcal{P}^{3N}$. The sum of $n$ components of $\vec{\chi}$ for quantum state $\rho$ can be expressed as
\begin{equation}
\mathrm{Tr}[(X_{n_1} +Y_{n_2} +Z_{n_3})\rho] \leq  \vec{\xi}^{\,\downarrow} \cdot \vec{\lambda}_{\rho}^{\downarrow} =\tau_n(X_{n_1}, Y_{n_2}, Z_{n_3}) \; , \label{n-comp-pure}
\end{equation}
where $n_1+n_2+n_3=n$; $\vec{\xi}$ is the eigenvalue list of $X_{n_1} +Y_{n_2} +Z_{n_3}$, and $\xi_1^{\downarrow}$ gives the maximum value of $\tau_n$. According to equation (\ref{Xn-set}), $X_{n_1}$ (similarly for $Y_{n_2}$ and $Z_{n_3}$) has $C(N,n_1)$ different choices, hence $\tau_n$ varies with the choices of $X_{n_1}$, $Y_{n_2}$, and $Z_{n_3}$,
\begin{equation}
\left\{\tau_n(X_{n_1}, Y_{n_2}, Z_{n_3})| X_{n_1} \in \mathcal{S}^{(x)}_{n_1}, Y_{n_2} \in \mathcal{S}^{(y)}_{n_2}, Z_{n_3} \in \mathcal{S}^{(z)}_{n_3}, \sum_{i=1}^3 n_i=n \right\}\; .
\end{equation}
Let $\vec{s}^{\,(n)} \in  \{\vec{\chi}^{\,\downarrow}\}$ be the vector that has the largest sum of the first $n$ components of various $\vec{\chi}^{\downarrow}$, then $\sum_{\mu=1}^{n} s_{\mu}^{(n)} = \max_{n_1,n_2,n_3} \{\tau_n \}$ where the maximization runs over different $n_i$ for $C(N,n_i)$ choices of $X_{n_1}$, $Y_{n_2}$, and $Z_{n_3}$. Note, for a given $n$, $\vec{s}^{\,(n)}$ may not be unique, but they all attribute equally to the vector $\vec{s}$ \cite{Maj-Latt-1}, and hence does not matter to our discussion. We then have the following optimal universal uncertainty relation, the main result of this work:
\begin{theorem}
In $N$-dimensional quantum system $\rho$, the probability distributions of measurements on $X$, $Y$, and $Z$ satisfy the following relation:
\begin{equation}
\vec{p} \oplus \vec{q} \oplus \vec{r} \prec \vec{s}\; . \label{Single-Maj-two}
\end{equation}
Here $\vec{s} := \vec{s}^{\,(1)} \vee \vec{s}^{\,(2)} \vee \cdots \vee \vec{s}^{\,(3N-1)}$ is the unique least upper bound of $\vec{p}\oplus \vec{q} \oplus \vec{r}$ over all quantum states. \label{Theorem-single-two}
\end{theorem}
\noindent {\bf Proof:} First to show $\vec{s}$ is an upper bound. From the definition of $\vec{s}$ and the associative law of operation $\vee$ on lattice, we have
\begin{equation}
\vec{s}^{\,(n)} \prec \vec{s} \; , \; \forall n\in \{1,\cdots, 3N-1\}\; .
\end{equation}
Since $\vec{s}^{\,(n)}$ owns the largest summed value of the first $n$ components in set $\{\vec{\chi}^{\,\downarrow}\}$, $\vec{p} \oplus \vec{q} \oplus \vec{r} \prec \vec{s}$ is satisfied by every quantum state.

Next, $\vec{s}$ is the least. For arbitrary $\vec{t}$, if $\vec{p} \oplus \vec{q} \oplus \vec{r} \prec \vec{t}$ for all quantum states, so is $\vec{s}^{\,(n)} \prec \vec{t},\; \forall n\in \{1,\cdots, 3N-1\}$, and according to Lemma \ref{Lemma-1}
\begin{equation}
\left. \begin{array}{l}
\vec{s}^{\,(1)} \prec \vec{t} \\ \vec{s}^{\,(2)} \prec \vec{t}
\end{array} \right\}  \Rightarrow \vec{s}^{\,(1)} \vee \vec{s}^{\,(2)} \prec \vec{t}\;. \label{Theorem-proof-equ-vee}
\end{equation}
Repeatedly applying equation (\ref{Theorem-proof-equ-vee}) to $\vec{s}^{\,(n)}$ will in the end lead to $\vec{s} \prec \vec{t}$. Q.E.D.

Note that the number of observables can be arbitrary in Theorem \ref{Theorem-single-two}, and the general POVM measurement also applies here. Moreover, the Theorem \ref{Theorem-single-two} is also applicable to mixed state with given $\lambda_{\rho}^{\downarrow}$ according to equation (\ref{n-comp-pure}), and $\vec{s}$ is optimal for such mixed state by maximizing the corresponding $\tau_n$. For Shannon entropy of $H(\vec{p}\,) := -\sum_{i}p_i\log p_i$, direct application of Theorem \ref{Theorem-single-two} leads to the following entropic uncertainty relation:
\begin{corollary}
For $M$ observables $X_{j}$, $j\in \{1, \ldots, M\}$, there exists the following entropic uncertainty relation
\begin{equation}
\sum_{j=1}^M H(X_{j}) \geq H(\vec{s}\,)  \; . \label{equation-single-two-entropy}
\end{equation}
Here $H(X_{j})= H(\vec{p}^{\,(j)})$ with $\vec{p}^{\,(j)}$ being the probability distribution of the measurement of $j$-th observable $X_{j}$; $\vec{s}$ is defined in Theorem \ref{Theorem-single-two} satisfying $\bigoplus \limits_{j=1}^M \vec{p}^{\,(j)} \prec \vec{s}$.
\label{Corollary-single-two-entropy}
\end{corollary}
Given that one has noticed the Shannon entropy is a Schur-concave function \cite{Majorization-Book}, the prove of equation (\ref{equation-single-two-entropy}) is quite straightforward. The Corollary 1 in fact can be further improved by adding a state-dependent term, i.e., from Theorem 3 of \cite{Entropy-improved-D}
\begin{equation}
\sum_{j=1}^M H(X_{j}) \geq H(\vec{s}\,) + D(\vec{s}\, \| \vec{\chi}\,) \; , \label{Improved-ent}
\end{equation}
where $\vec{\chi} = \bigoplus \limits_{j=1}^M \vec{p}^{\,(j)} $ and $D(\vec{s}\,||\vec{\chi}) \equiv \sum \limits_j s_j^{\downarrow} \log(\frac{s_j^{\downarrow}}{\chi_j^{\downarrow}})$. As the relative entropy is nonnegative, it can be easily verified that $D(\vec{s}\,||\vec{\chi}) = MD(\frac{1}{M}\vec{s}\,|| \frac{1}{M} \vec{\chi})\geq 0$.

For a given set of incompatible observables, e.g. $X$, $Y$, and $Z$, quantum states $\rho_1$ and $\rho_2$ will result in two probability vectors $\vec{\chi}_1, \vec{\chi}_2 \in \mathcal{P}^{3N}$, where without loss of generality we have assumed the components of $\vec{\chi}_{1,2}$ are arranged in non-increasing order. The property of the majorization lattice tells that there exists a distance measure on $\mathcal{P}^{3N}$ \cite{Maj-Latt-3}
\begin{equation}
d(\vec{\chi}_1, \vec{\chi}_2) := H(\vec{\chi}_1) + H(\vec{\chi}_2) - 2H(\vec{\chi}_1 \vee \vec{\chi}_2) \geq 0\; . \label{Metric-dist}
\end{equation}
In account of this metric, we may get the following corollary:
\begin{corollary}
For arbitrary different probability distribution vectors $\vec{\chi}_1$ and $\vec{\chi}_2$, we have the entropic uncertainty relation
\begin{align}
H(\vec{\chi}_1) + H(\vec{\chi}_2) \geq 2H(\vec{s}\,) + d(\vec{\chi}_1,\vec{\chi}_2) \;. \label{Coro-equ}
\end{align}
The $d(\vec{\chi}_1,\vec{\chi}_2) >0$ while $\vec{\chi}_1$ and $\vec{\chi}_2$ are different vectors.
\label{Corollary-2}
\end{corollary}
\noindent{\bf Proof:} The lattice theory tells that, if $\vec{\chi}_i \prec \vec{s}$, then $\vec{\chi}_i \vee \vec{s} = \vec{s}$ for both $i=1,2$, and hence
\begin{equation}
d(\vec{\chi}_i,\vec{s}\,) = H(\vec{\chi}_i) - H(\vec{s}\,) \; .
\end{equation}
Since $d(\vec{\chi}_1, \vec{\chi}_2) \leq d(\vec{\chi}_1,\vec{s}\,) +d(\vec{\chi}_2,\vec{s}\,)$ \cite{Maj-Latt-3}, equation (\ref{Coro-equ}) is readily obtained. Q.E.D.

Corollary \ref{Corollary-2} reveals an important feature of majorization lattice, that is the sum of the two independent uncertainty relations $H(\vec{\chi}_{1,2}) \geq  H(\vec{s}\,)$ yields a even stronger one due to $d(\vec{\chi}_1,\vec{\chi}_2) \geq 0$. The distance measure $d(\cdot,\cdot)$ induces a metric on $\mathcal{P}^{3N}$ \cite{Maj-Latt-3} and is non-zero even if $H(\vec{\chi}_1)=H(\vec{\chi}_2)$. As there exist the distributions that neither $\vec{\chi}_1 \prec \vec{\chi}_2$ nor $\vec{\chi}_1 \succ \vec{\chi}_2$, we call such distribution vectors incomparable, which means that they have no particular order in majorization relation. The time-order-event analogy may be heuristic in the understanding of incomparability, that is, two space-like separated events have no particular order in time. Considering that entropies are always comparable, say either $H(\vec{\chi}_1) < H(\vec{\chi}_2)$ or $H(\vec{\chi}_1) \geq H(\vec{\chi}_2)$ (also true for variances), the variance and entropy may be called the ``absolute measure'' of quantum uncertainty, while the lattice theory reveals the intrinsic structure of quantum uncertainty and leads merely to ``relative measure''.

It is also interesting to compare our results with that of \cite{Majorization-1} and \cite{Majorization-2}. Note, there is no explicit definition for the least upper bound of the majorization uncertainty relation in either of \cite{Majorization-1} and \cite{Majorization-2}. The algorithm used in \cite{Majorization-1} and \cite{Majorization-2} for getting the upper bound of the majorization relation goes as follows: First compute $\Omega_k$ which has the largest value of the sum of the first $k$ components on the left hand side of the majorization relation, then regard $\vec{t} = (\Omega_1, \Omega_2-\Omega_1,\Omega_3-\Omega_2 , \cdots )^{\mathrm{T}}$ as the upper bound for the majorization uncertainty relation. It can be shown that this algorithm can not guarantee obtaining the least upper bound in general. For example, the example 1 in \cite{Maj-Latt-1}, for two probability distributions
\begin{align}
\vec{p} = (0.6,0.15,0.15,0.1)\; , \; \vec{q} = (0.5,0.25,0.20,0.05) \; ,
\end{align}
it is easy to know that the $\Omega_k$ for $\vec{p}$ and $\vec{q}$ are $\Omega_1=0.6$, $\Omega_2=0.75$, $\Omega_3=0.95$, $\Omega_4=1$.
However the vector
\begin{align}
\vec{t} = (\Omega_1,\Omega_2-\Omega_1,\Omega_3-\Omega_2,\Omega_4-\Omega_3) = (0.6,0.15,0.2,0.05) \;
\end{align}
is not the least upper bound for $\vec{p},\vec{q}\prec \vec{t}$ noticing
\begin{align}
\vec{p},\vec{q} \prec (0.6,0.175,0.175,0.05) \prec \vec{t} \; .
\end{align}
In fact in this case the least upper bound is $\vec{s} = (0.6,0.175,0.175,0.05)$. In all, the upper bound obtained by means of finding the maximum value of the sum of the first largest $k$ components of a probability vector in a set will not always give the least upper bound for the set of probability vectors in the majorization relation, and a systematic flattening operation is generally needed (see Lemma 3 in \cite{Maj-Latt-1}).

\subsection{The optimality of the uncertainty relation and Lorenz curve}

To elucidate the physics embedded in above mathematics, following we give three typical examples for illustration, but the calculation details will be given in Appendix.

\begin{figure}[t]\centering
\scalebox{0.28}{\includegraphics{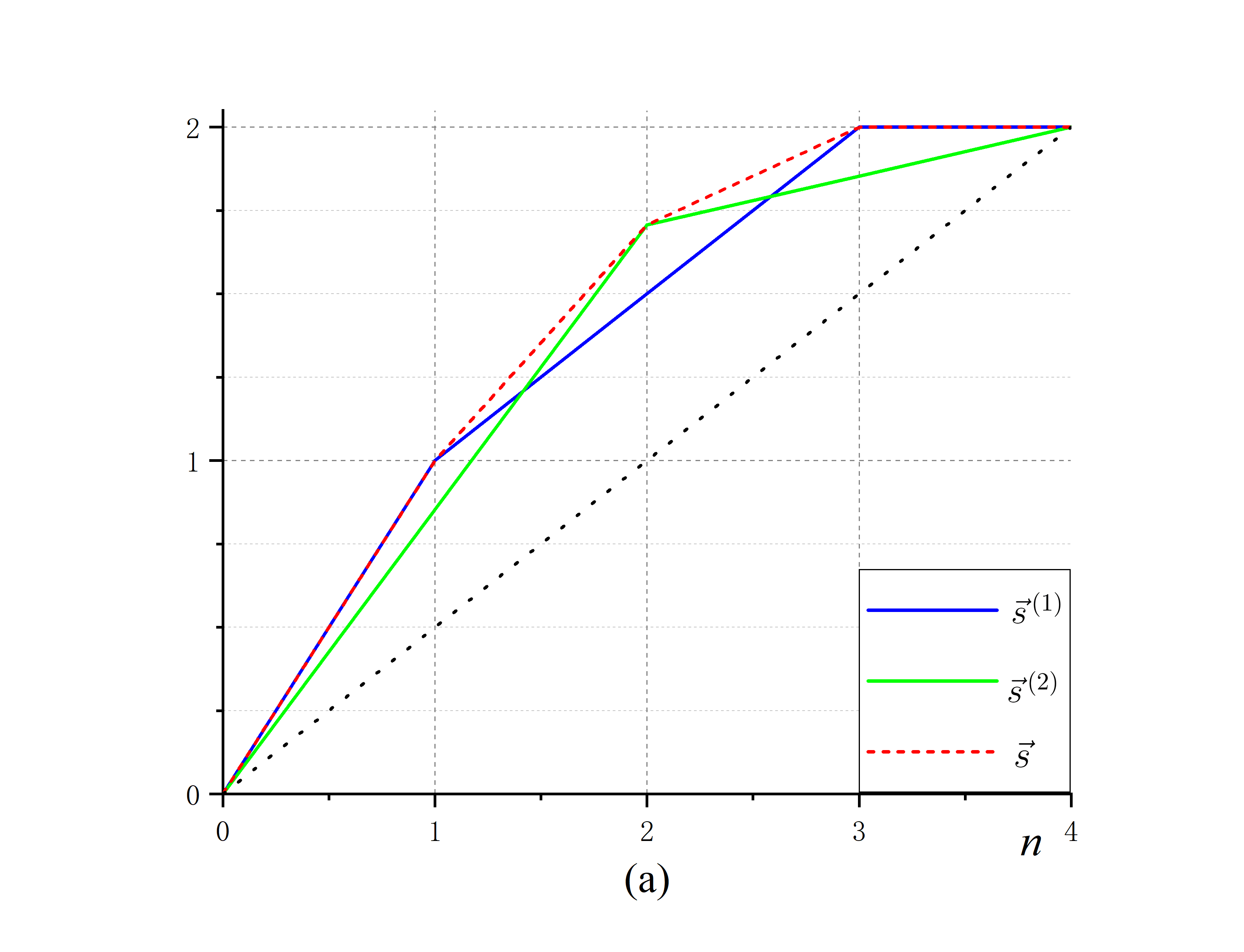}}
\scalebox{0.28}{\includegraphics{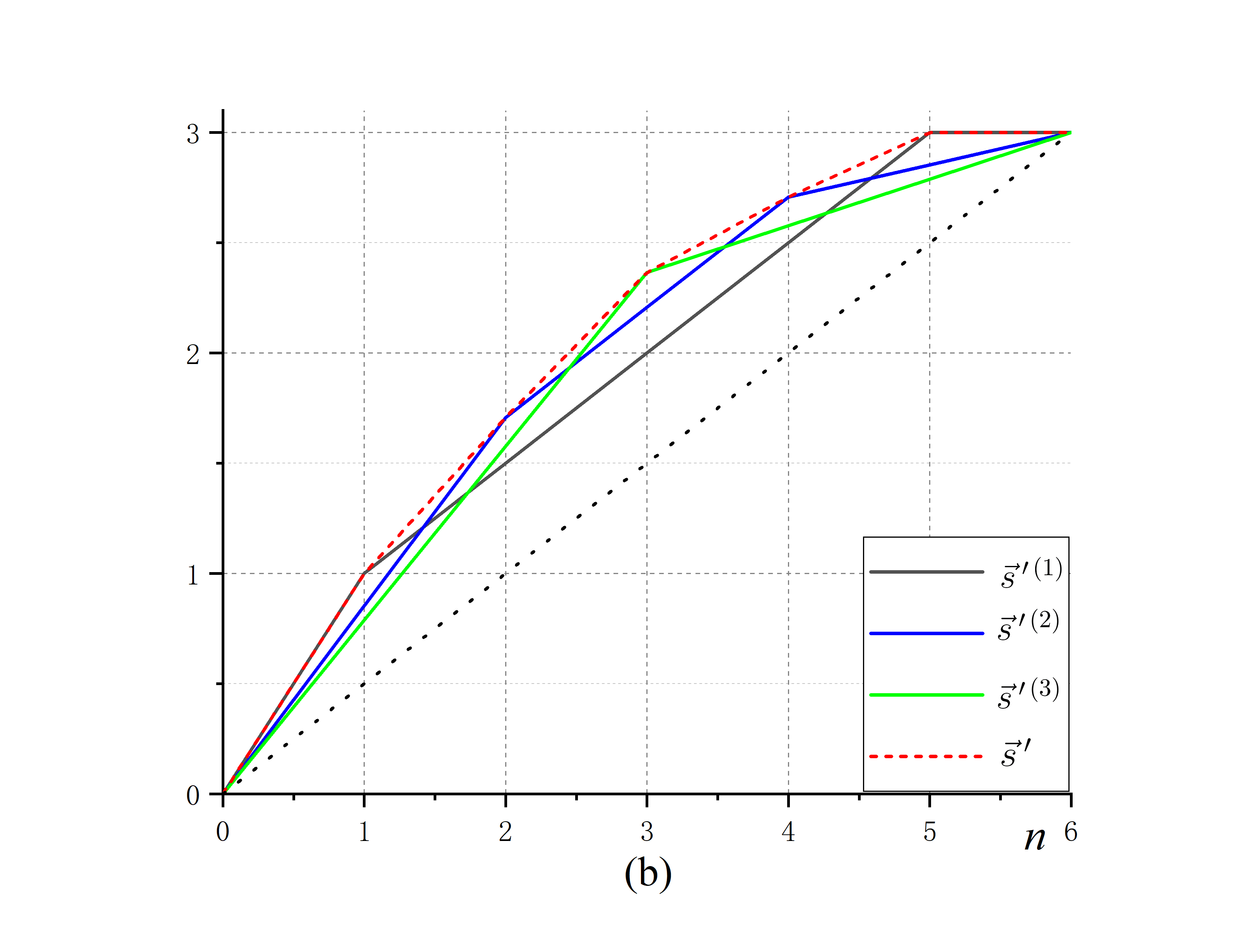}}
\caption{The Lorenz curves for the universal uncertainty relations of two and three observables. (a) the Lorenz curves for observables $X$ and $Z$ where $\vec{\chi} = \vec{p}_x \oplus \vec{p}_z$; (b) the Lorenz curves for observables $X$, $Y$, and $Z$ where $\vec{\chi}' = \vec{p}_x \oplus \vec{p}_y \oplus \vec{p}_z$. By means of $\vec{s} = \vec{s}^{\,(1)} \vee  \vec{s}^{\,(2)}$ and  $\vec{s}\,' =  \vec{s}\,'^{(1)} \vee  \vec{s}\,'^{(2)} \vee \vec{s}\,'^{(3)}$, the Lorenz curves of $\vec{s}$ and $\vec{s}\,'$ (red dashed lines) give the least possible envelops enclosing the curves of $\vec{\chi}$ and $\vec{\chi}'$ for all quantum states. } \label{Figure-1}
\end{figure}

Example one: considering the following two observables in the general qubit system
\begin{equation}
Z = \begin{pmatrix}
1 & 0 \\
0 & -1
\end{pmatrix}\; , \;
X = \begin{pmatrix}
\cos\theta & \sin\theta\\
\sin\theta & -\cos\theta
\end{pmatrix}\;, \; \theta\in [0,\frac{\pi}{2}]\; , \label{Example-qubit-xz}
\end{equation}
the probability distribution vectors $\vec{\chi}=\vec{p}_x\oplus \vec{p}_z$ are then four dimensional, and
\begin{align}
\vec{s}^{\,(1)} & = (\lambda_1, \lambda_1\cos^2\frac{\theta}{2}+\lambda_2\sin^2\frac{\theta}{2}, \lambda_1\sin^2\frac{\theta}{2}+ \lambda_2\cos^2\frac{\theta}{2}, \lambda_2)^{\mathrm{T}} =\vec{s}^{\,(3)}\; , \\
\vec{s}^{\,(2)} & = (\lambda_1 \cos^2\frac{\theta}{4}+ \lambda_2 \sin^2\frac{\theta}{4}, \lambda_1 \cos^2\frac{\theta}{4}+ \lambda_2 \sin^2\frac{\theta}{4}, \nonumber \\
& \hspace{0.7cm}\lambda_1 \sin^2\frac{\theta}{4} + \lambda_2 \cos^2\frac{\theta}{4}, \lambda_1 \sin^2\frac{\theta}{4} + \lambda_2 \cos^2\frac{\theta}{4})^{\mathrm{T}}\; .
\end{align}
Following the procedure given in Ref. \cite{Maj-Latt-1}, we have
\begin{align}
\vec{s} & = \vec{s}^{\,(1)} \vee \vec{s}^{\,(2)} \nonumber \\
& = (\lambda_1, \lambda_1\cos\frac{\theta}{2} + 2\lambda_2 \sin^2 \frac{\theta}{4}, 2\lambda_1 \sin^2\frac{\theta}{4} +\lambda_2\cos\frac{\theta}{2}, \lambda_2)^{\mathrm{T}} \; .
\end{align}
We find the probability distribution vectors $\vec{s}^{\,(1)}$, $\vec{s}^{\,(2)}$, and $\vec{s}$ may be exhibited by the Lorenz curve, as shown in Figure \ref{Figure-1}(a) for $\lambda_1=1$ and $\theta=\frac{\pi}{2}$. The Lorenz curve of a probability distribution vector $\vec{\chi}$ is $y_{\chi}:= f_{\chi}(n) = \sum_{i=1}^n \chi_i^{\downarrow}$ with $f_{\chi}(0)=0$.

For completely mixed state $\rho = \frac{1}{2}\mathds{1}$, the probability distribution $\vec{\chi}_{\mathrm{mix}} = (\frac{1}{2}, \frac{1}{2}, \frac{1}{2}, \frac{1}{2})^{\mathrm{T}}$, whose Lorenz curve goes from (0,0) to (4,2), the dashed anti-diagonal line in Figure \ref{Figure-1}(a). The Lorenz curve of each $\vec{\chi}$ lies below the curve of $\vec{s}$ and above the anti-diagonal line $\vec{\chi}_{\mathrm{mix}}$. Obviously, the Lorenz curve of $\vec{s}$ is the least possible envelope, red dashed line in Figure \ref{Figure-1}(a), enclosing the curves of $\vec{s}^{\,(n)}$, and is optimal for the universal uncertainty relation $\vec{p}_x\oplus \vec{p}_z \prec \vec{s}$ for any quantum states.

Example two: for $X=\sigma_x$, $Y=\sigma_y$, and $Z=\sigma_z$ in pure qubit system, we can find the optimal bound for $\vec{p}_x \oplus \vec{p}_y \oplus \vec{p}_z \prec \vec{s}\,'$. The vectors $\vec{s}\,'^{(n)}$, which have the largest sum of the first $n$ components, are
\begin{align}
\vec{s}\,'^{(1)} & = (1,\frac{1}{2}, \frac{1}{2},\frac{1}{2}, \frac{1}{2},0) = \vec{s}\,'^{(5)} \; , \\
\vec{s}\,'^{(2)} & = (\frac{1}{2-\sqrt{2}},\frac{1}{2-\sqrt{2}}, \frac{1}{2}, \frac{1}{2}, \frac{1}{2+\sqrt{2}}, \frac{1}{2+\sqrt{2}}) = \vec{s}\,'^{(4)}\; , \\
\vec{s}\,'^{(3)} & = (\frac{1}{3-\sqrt{3}}, \frac{1}{3-\sqrt{3}}, \frac{1}{3-\sqrt{3}}, \frac{1}{3+\sqrt{3}}, \frac{1}{3+\sqrt{3}}, \frac{1}{3+\sqrt{3}}) \; .
\end{align}
The $\vec{s}\,'$ can then be readily obtained
\begin{align}
\vec{s}\,' & = \vec{s}\,'^{(1)} \vee \vec{s}\,'^{(2)} \vee \vec{s}\,'^{(3)} \nonumber \\
& = (1, \frac{\sqrt{2}}{2}, \frac{1+\sqrt{3}-\sqrt{2}}{2},\frac{1-\sqrt{3}+\sqrt{2}}{2}, \frac{2-\sqrt{2}}{2},0 )^{\mathrm{T}} \; .
\end{align}
The Lorenz curves of $\vec{s}\,'$ and $\vec{s}\,'^{(i)}$ are plotted in Figure \ref{Figure-1}(b), where the optimality of $\vec{s}\,'$ for universal uncertainty relation is evidently demonstrated.

Example three: for 3-dimensional observables $X$ and $Y$ with the orthonormal bases of Ref.\cite{Example-Coles-3}, i.e.,
\begin{align}
(|x_1\rangle, |x_2\rangle, |x_3\rangle) &= \begin{pmatrix}
1 & 0 & 0 \\
0 & 1 & 0 \\
0 & 0 & 1
\end{pmatrix} \;, \\
 (|y_1\rangle, |y_2\rangle, |y_3\rangle) &=
\begin{pmatrix}
\frac{1}{\sqrt{3}} & \frac{1}{\sqrt{3}} & \frac{1}{\sqrt{3}} \\
\frac{1}{\sqrt{2}} & 0 & -\frac{1}{\sqrt{2}} \\
\frac{1}{\sqrt{6}} & -\sqrt{\frac{2}{3}} & \frac{1}{\sqrt{6}}
\end{pmatrix} \; ,
\end{align}
the optimal bound for universal uncertainty relation reads
\begin{equation}
\vec{p}_x \oplus \vec{p}_y \prec \vec{s}\,''= (1,\frac{\sqrt{6}}{3}, 1-\frac{\sqrt{6}}{3},0,0,0)\; .
\end{equation}
Here $\vec{s}\,''$ is optimal in the sense that for any vector $\vec{v}$ satisfying $\vec{p}_x \oplus \vec{p}_y \prec \vec{v}$ in all quantum states, $\vec{s}\,'' \prec \vec{v}$.

To summarize, the application of majorization lattice to the study of uncertainty principle gives rise to some new insights, that the sole scalar measure, such as variance or entropy, is insufficient to characterize the quantum uncertainty. The variance or entropy maps the probability distribution into a real number, while in fact the distribution uncertainty has some intrinsic structures as revealed by the majorization lattice. The structure of quantum uncertainty also deciphers the puzzle: why getting the optimal bound for the variance- or entropy-based uncertainty relation is a tough issue. Each scalar quantity can only measure one facet of the multifaceted quantum uncertainty and hence the bound may vary with the measures chosen.

For one type of entropic function, e.g., the Shannon entropy $H(\cdot)$, the majorization lattice can tell why there are hurdles in getting the optimal entropic bound. We know the procedure of optimizing entropic uncertainty relation is to find the minimum $H(\vec{\chi})$ over all quantum states. In order to get the minimum value, the vector $\vec{\chi}_{\mathrm{min}}$ should be incomparable with $\vec{s}^{\,(n)}$ under the majorization relation, that is the Lorenz curves of $\vec{s}^{\,(n)}$ intercross with that of $\vec{\chi}_{\mathrm{min}}$. For incomparable vectors under majorization, there exists the catalytic phenomenon initially observed in entanglement transformation under local quantum operations and classical communication \cite{Ent-assisted}, which causes the comparison of different entropic measures complicated. That is, for $\vec{\chi}_{\mathrm{min}} \nprec \vec{s}^{\,(n)}$ and $\vec{s}^{\,(n)} \nprec \vec{\chi}_{\mathrm{min}}$, there may exist an unknown catalytic probability tensor that predetermines the relative size of $H(\vec{\chi}_{\mathrm{min}})$ and $H(\vec{s}\,)$ \cite{IEEE-assisted}. The optimization of entropic uncertainty relation is then turned to finding the quantum state whose $\vec{\chi}$ catalytically majorizes others, which is hard to be solved analytically \cite{Ent-assisted}. It is worth mentioning that majorization lattice has, and may have more, profound applications in the entanglement transformation \cite{Maj-latt-ent-1, Maj-latt-ent-2}.

It is worth mentioning that with different entropic functions, one may even get contradicting results. For example, two probability distributions $\vec{p}_1=(\frac{1}{2},\frac{1}{2},0)$ and $\vec{p}_2=(\frac{1}{12},\frac{1}{12},\frac{5}{6})$ may lead to
\begin{align}
H_{\frac{1}{5}}(\vec{p}_1) > H_{\frac{1}{5}}(\vec{p}_2)\;\ \&\ \; H_{2}(\vec{p}_1) < H_{2}(\vec{p}_2) \; ,
\end{align}
where the R\'enyi entropies $H_{\alpha}(\vec{v}\,) := \frac{1}{1-\alpha}\log(\sum_{i}v_i^{\alpha})$ with different index $\alpha$ are scalar measures of uncertainty. It should be noted that when the number of observables and the dimensions of the system go large, computer programs may be used to simplify the solving of the least upper bound of the majorization relation, i.e., semidefinite programming \cite{SDP}.

\section{Conclusions}

In this work we have explored the uncertainty relation by employing the lattice theory, and obtained the optimal bound for universal uncertainty relation, which is applicable to general measurement and arbitrary number of observables. The application of lattice theory indicates that the quantum uncertainty is a structure quantity, the variance or entropy may not be the most appropriate measure for it. Moreover, we find the optimality of the uncertainty relation can be intuitively exhibited by the Lorenz curve, which enables the direct experimental test of universal uncertainty relation and may even shed some light on the understanding of economic phenomena. Finally, the majorization lattice reveals the incomparability of quantum uncertainties, that does not manifest in variance or entropy form. This character indicates that the optimization of entropic or variance-based uncertainty relation generally must be a tough issue.

\section*{Acknowledgements}
\noindent
This work was supported in part by the Ministry of Science and Technology of the Peoples' Republic of China(2015CB856703); by the Strategic Priority Research Program of the Chinese Academy of Sciences, Grant No.XDB23030100; and by the National Natural Science Foundation of China(NSFC) under the Grants 11375200 and 11635009.

\newpage

\setcounter{figure}{0}
\renewcommand{\thefigure}{S\arabic{figure}}
\setcounter{equation}{0}
\renewcommand\theequation{S\arabic{equation}}
\setcounter{theorem}{0}
\renewcommand{\thetheorem}{S\arabic{theorem}}
\setcounter{observation}{0}
\renewcommand{\theobservation}{S\arabic{observation}}
\setcounter{proposition}{0}
\renewcommand{\theproposition}{S\arabic{proposition}}
\setcounter{lemma}{0}
\renewcommand{\thelemma}{S\arabic{lemma}}
\setcounter{corollary}{0}
\renewcommand{\thecorollary}{S\arabic{corollary}}
\setcounter{section}{0}
\renewcommand{\thesection}{S\arabic{section}}

\appendix{\bf \LARGE Appendix}

For the sake of integrity, here we present some basic properties of majorization lattice and the method for constructing the least upper bound for the majorization lattice in Section A. Section B contains the detailed derivations for the examples of qubit and qutrit states.

\section{The majorization lattice}

\subsection{Basic defintions}

The majorization relation between two tuples of real numbers is defined as \cite{Supp-Majorization-Book}:
\begin{equation}
\vec{p} \prec \vec{q} \; \Longleftrightarrow \; \sum_{i=1}^k {p}^{\downarrow}_i \leq \sum_{j=1}^k {q}^{\downarrow}_j \; , \; k\in \{1,\ldots, N\} \; ,
\end{equation}
where the superscript $\downarrow$ means that the components of vectors $\vec{p}$ and $\vec{q}$ are arrayed in descending order, and the equality holds when $k=N$. Let $\mathcal{P}^{N}$ be the set of all $N$-dimensional probability distributions with components in nonincreasing order
\begin{equation}
\mathcal{P}^{N} = \left\{ \vec{p} = (p_1,\cdots, p_N)^{\mathrm{T}}\left| p_i \in [0,1] \; , \; \sum_{i=1}^N p_i = \mathrm{const.}\; , \; p_i\geq p_{i+1} \right. \right\} \; .
\end{equation}
The quadruple $\langle \mathcal{P}^N, \prec, \wedge, \vee \rangle$ form a lattice, where $\mathcal{P}^N$ is a set, $\prec$ is a partial ordering on $\mathcal{P}^N$, and there is a unique greatest lower bound $\vec{p} \wedge \vec{q}$ (meet) and a unique least upper bound $\vec{p} \,\vee \vec{q}$ (join). The demonstration  that $\mathcal{P}^{N}$ is a lattice can be found in  \cite{Supp-AU-Book, Supp-RB-Bapat, Supp-Comm, Supp-Maj-Latt-1}.

\subsection{Construction of the least upper bound $\vec{p} \vee \vec{q}$}

The construction of $\vec{p} \vee \vec{q}$ for $\vec{p}, \vec{q} \in \mathcal{P}^N$ can be found in \cite{Supp-Maj-Latt-1}. Here we summarize their procedure as follows.

First, we define the vector $\beta(\vec{p},\vec{q}\,)$ whose components are $b_i$ and
\begin{align}
b_i & = \max\left\{ \sum_{j=1}^i p_j, \sum_{j=1}^i q_j \right\} - \sum_{j=1}^{i-1} b_j \nonumber \\
& = \max\left\{ \sum_{j=1}^i p_j, \sum_{j=1}^i q_j \right\} - \max\left\{ \sum_{j=1}^{i-1} p_j, \sum_{j=1}^{i-1} q_j \right\} \; .
\end{align}
While $\beta(\vec{p},\vec{q}\,)^{\downarrow} \in \mathcal{P}^N$, $\beta(\vec{p},\vec{q}\,)$ may not be in the set $\mathcal{P}^N$.

Second, there exists the following Lemma (Lemma 3 of \cite{Supp-Maj-Latt-1})
\begin{lemma}
Let $\beta(\vec{p},\vec{q}\,) = (b_1,\cdots,b_N)^{\mathrm{T}}$, and let $j$ be the smallest integer in $\{2,\cdots, N\}$ such that $b_j>b_{j-1}$. Moreover, let $i$ be the greatest integer in $\{1,2,\cdots, j-1\}$ such that
\begin{align}
b_{i-1} \geq \frac{\sum_{r=i}^{j} b_r}{j-i+1} = a \; .
\end{align}
Let the probability distribution $\vec{\mu} = (\mu_1,\cdots, \mu_{N})$ be defined as
\begin{align}
\mu_{r} = \left\{
\begin{array}{ll}
a & \mathrm{for}\, r=i, i+1,\cdots, j\\
b_r & \mathrm{otherwise.}
\end{array}\right.
\end{align}
Then for the probability distribution $\vec{\mu}$ we have that
\begin{align}
\mu_{r-1} \geq \mu_r \; , \forall r=2,\cdots, j \;
\end{align}
and
\begin{align}
\sum_{s=1}^k \mu_{s} \geq \sum_{s=1}^k b_s \; , k=1,\cdots, N \;.
\end{align}
Moreover, for all $\vec{t}=(t_1,\cdots, t_N) \in \mathcal{P}^{N}$ such that
\begin{align}
\sum_{s=1}^k t_s \geq \sum_{s=1}^k b_s\; , k=1,\cdots, N
\end{align}
we also have
\begin{align}
\sum_{s=1}^kt_s \geq \sum_{s=1}^k \mu_s \;, k=1,\cdots, N\;.
\end{align}
\label{supp-lemma1}
\end{lemma}

Finally, if $\beta(\vec{p}, \vec{q}\,) \in \mathcal{P}^N$, i.e., there is no $j$ such that $b_j>b_{j-1}$, then $\beta(\vec{p}, \vec{q}\,) = \vec{p} \vee \vec{q}$. If $\beta(\vec{p}, \vec{q}\,) \notin \mathcal{P}^N$, by iteratively applying the transformation described in Lemma \ref{supp-lemma1} with no more than $N-1$ iterations, we eventually obtain a vector $\vec{s} \in \mathcal{P}^N$ such that, $\vec{p}, \vec{q} \prec \vec{s}$, and for any vector $\vec{t} \in \mathcal{P}^{N}$ such that $\vec{p} \prec \vec{t}$ and $\vec{q} \prec \vec{t}$, it holds also that $\vec{s} \prec \vec{t}$. And therefore $\vec{s} = \vec{p} \vee \vec{q}$.

In order to construct the least upper bound for more than two probability distribution vectors we need the following theorem for a lattice (Theorem 2.9 in \cite{Supp-DP-book})
\begin{theorem}
Let $\langle \mathcal{P}^N, \prec, \wedge, \vee \rangle$ be a lattice. Then $\vee$ and $\wedge$ satisfy, for all $\vec{a}, \vec{b},\vec{c} \in \mathcal{P}^N$
\begin{align}
(\vec{a}\vee \vec{b}\,) \vee \vec{c} = \vec{a}\vee (\vec{b} \vee \vec{c}\,)\; , \;  (\vec{a} \wedge \vec{b}\,) \wedge \vec{c} = \vec{a} \wedge (\vec{b} \wedge \vec{c}\,)\; ,  \nonumber \\
 \vec{a} \vee \vec{b} = \vec{b} \vee \vec{a}\; , \;  \vec{a} \wedge \vec{b}  = \vec{b} \wedge \vec{a}\; , \; \vec{a} \vee \vec{a} = \vec{a}\; ,\;
\vec{a} \wedge \vec{a} = \vec{a}\; , \nonumber \\
\vec{a} \vee (\vec{a} \wedge \vec{b}\, ) = \vec{a} \; , \; \vec{a} \wedge (\vec{a} \vee \vec{b}\,) = \vec{a} \; .
\end{align}
\end{theorem}
In a lattice, associativity of join $\vee$ and meet $\wedge$ allows us to write iterated joins and meets unambiguously.

\section{Examples of qubit and qutrit states}

\subsection{Two observables in general qubit system}

For the two observables in qubit system
\begin{equation}
Z = \begin{pmatrix}
1 & 0 \\
0 & -1
\end{pmatrix}\; , \;
X = \begin{pmatrix}
\cos\theta & \sin\theta\\
\sin\theta & -\cos\theta
\end{pmatrix}\;, \; \theta\in [0,\frac{\pi}{2}]\; , \label{Example-qubit-xz}
\end{equation}
the projective measurement bases of $Z$ and $X$ are
\begin{align}
u_z = (|z_1\rangle, |z_2\rangle) =
\begin{pmatrix}
1 & 0 \\
0 & 1
\end{pmatrix}\; , \;  u_x = (|x_1\rangle, |x_2\rangle) =
\begin{pmatrix}
\cos\frac{\theta}{2} & \sin\frac{\theta}{2} \\
\sin\frac{\theta}{2} & -\cos\frac{\theta}{2}
\end{pmatrix}\; ,
\end{align}
where $Z=|z_1\rangle\langle z_1| -|z_2\rangle \langle z_2|$ and $X=|x_1\rangle\langle x_1| -|x_2\rangle \langle x_2|$. The probability distribution vectors $\vec{\chi}=\vec{p}_x\oplus \vec{p}_z$ are then four dimensional. Our aim is finding the least bound $\vec{s}$ where $\vec{p}_x\oplus \vec{p}_z \prec \vec{s}$ is satisfied for all quantum states.

First, for quantum states with $\vec{\lambda}_{\rho}^{\downarrow} = (\lambda_1,\lambda_{2})$, if we want $\vec{\chi}$ has one largest component, we need to maximize $\mathrm{Tr}[O_i \rho]$ where
\begin{align}
O_1 = |x_1\rangle \langle x_1| \; , \; O_2 = |x_1\rangle \langle x_1| \; , \; O_3 = |z_1\rangle \langle z_1| \; , \; O_4 = |z_2\rangle \langle z_2| \; .
\end{align}
It is easy to observe that the maximal value is $\max\{\tau_1\} = \lambda_1$ which may be obtained, for instance, by $\rho_1 = \lambda_1|x_1\rangle \langle x_1| + \lambda_2 |x_2\rangle \langle x_2|$, $\rho_2 = \lambda_1|z_1\rangle \langle z_1| + \lambda_2 |z_2\rangle \langle z_2|$, etc. Taking $\rho_1$ into $\vec{\chi}=\vec{p}_x\oplus \vec{p}_z$, we have
\begin{equation}
\vec{s}^{\,(1)} = (\lambda_1, \lambda_1\cos^2\frac{\theta}{2}+\lambda_2\sin^2\frac{\theta}{2}, \lambda_1\sin^2\frac{\theta}{2}+ \lambda_2\cos^2\frac{\theta}{2}, \lambda_2)^{\mathrm{T}} \; .
\end{equation}
Here the components have been rearranged in descending order. Since $\vec{s}^{\,(1)}$ also has the largest sum of any $3$ components, we have $\vec{s}^{\,(3)} = \vec{s}^{\,(1)}$.

Second, in order to get the largest sum of any two components of $\vec{\chi}$, we need to maximize $\mathrm{Tr}[O_i\rho]$ where
\begin{align}
& O_1 = |x_1\rangle\langle x_1| + |x_2\rangle\langle x_2| \; , \; O_2 =|x_1\rangle\langle x_1| + |z_1\rangle\langle z_1| \; , \; O_3 = |x_1\rangle\langle x_1| + |z_2\rangle\langle z_2| \; , \\
& O_4 =|x_2\rangle \langle x_2| + |z_1\rangle \langle z_1| \; , \; O_5 =|x_2\rangle \langle x_2| + |z_2\rangle \langle z_2| \; , \; O_6 =|z_2\rangle\langle z_1| + |z_2\rangle\langle z_1| \; .
\end{align}
The maximum value is simply $\max\{\tau_2\} = \max_{i}\{ \xi_1^{(i)} \lambda_1 + \xi_2^{(i)} \lambda_2\}$ with $\xi_{1,2}^{(i)}$ being the eigenvalues of $O_i$ in descending order. The probability vector $\vec{\chi}$ with the largest sum of any two components reads
\begin{align}
\vec{s}^{\,(2)} & = (\lambda_1 \cos^2\frac{\theta}{4}+ \lambda_2 \sin^2\frac{\theta}{4}, \lambda_1 \cos^2\frac{\theta}{4}+ \lambda_2 \sin^2\frac{\theta}{4}, \nonumber \\
& \hspace{0.7cm}\lambda_1 \sin^2\frac{\theta}{4} + \lambda_2 \cos^2\frac{\theta}{4}, \lambda_1 \sin^2\frac{\theta}{4} + \lambda_2 \cos^2\frac{\theta}{4})^{\mathrm{T}}\; ,
\end{align}
which can be obtained by $\rho = \lambda_1|\phi_+\rangle\langle \phi_+|+\lambda_2|\phi_-|\rangle \langle \phi_-|$. Here $|\phi_{+}\rangle = \cos\frac{\theta}{4} |z_1\rangle + \sin\frac{\theta}{4} |z_2\rangle$ and
$|\phi_{-}\rangle = -\sin\frac{\theta}{4} |z_1\rangle + \cos\frac{\theta}{4} |z_2\rangle$
are eigenvectors of $O_2$ whose eigenvalues are $2\cos^2\frac{\theta}{4}$ and $2\sin^2\frac{\theta}{4}$. It can be checked that $\max\{\tau_2\} = \xi_1^{(2)} \lambda_1 + \xi_2^{(2)} \lambda_2 = 2\lambda_1 \cos^2\frac{\theta}{4} + 2\lambda_2 \sin^2\frac{\theta}{4} $

Finally, the least upper bound of $\vec{p}_x\oplus \vec{p}_z \prec \vec{s}$ now can be obtained via the following
\begin{align}
\vec{s} & = \vec{s}^{\,(1)} \vee \vec{s}^{\,(2)} \nonumber \\
& = (\lambda_1, \lambda_1\cos\frac{\theta}{2} + 2\lambda_2 \sin^2 \frac{\theta}{4}, 2\lambda_1 \sin^2\frac{\theta}{4} +\lambda_2\cos\frac{\theta}{2}, \lambda_2)^{\mathrm{T}} \; .
\end{align}
Here $\vec{s}$ is the unique least upper bound of the uncertainty relation $\vec{p}_x\oplus \vec{p}_z \prec \vec{s}$ for all the quantum states whose eigenvalues are $\{\lambda_1, \lambda_2\}$.

\subsection{Three observables in pure qubit system}

For three observables of $X=\sigma_x$, $Y=\sigma_y$, and $Z=\sigma_z$ in pure qubit system, we can find the optimal bound for $\vec{p}_x \oplus \vec{p}_y \oplus \vec{p}_z \prec \vec{s}\,'$. The vectors $\vec{s}\,'^{(n)}$, which have the largest sum of first $n$ components, are
\begin{align}
\vec{s}\,'^{(1)} & = (1,\frac{1}{2}, \frac{1}{2},\frac{1}{2}, \frac{1}{2},0) = \vec{s}\,'^{(5)} \; , \\
\vec{s}\,'^{(2)} & = (\frac{1}{2-\sqrt{2}},\frac{1}{2-\sqrt{2}}, \frac{1}{2}, \frac{1}{2}, \frac{1}{2+\sqrt{2}}, \frac{1}{2+\sqrt{2}}) = \vec{s}\,'^{(4)}\; , \\
\vec{s}\,'^{(3)} & = (\frac{1}{3-\sqrt{3}}, \frac{1}{3-\sqrt{3}}, \frac{1}{3-\sqrt{3}}, \frac{1}{3+\sqrt{3}}, \frac{1}{3+\sqrt{3}}, \frac{1}{3+\sqrt{3}}) \; .
\end{align}
Let the eigenvectors of $X$, $Y$, and $Z$ be $\{|+\rangle, |-\rangle\}$, $\{|L\rangle, |R\rangle\}$, and $|H\rangle, |V\rangle$ respectively, then $\vec{s}\,^{(1)}$ can be obtained by any one of the eigenvectors of $X$, $Y$, and $Z$; $\vec{s}\,^{(2)}$ may be obtained by eigenvector of $|+\rangle\langle +| + |H\rangle \langle H|$ with the larger eigenvalue $\frac{2+\sqrt{2}}{2} = \max\{\tau_2\}$; $\vec{s}\,^{(3)}$ may be obtained by the eigenvector of $|+\rangle \langle +| + |L\rangle \langle L| + |H\rangle \langle H|$ with the larger eigenvalue $\frac{3+\sqrt{3}}{2} = \max\{\tau_3\}$. We have
\begin{align}
|\psi^{(1)}\rangle & = (1,0) \; ,\; \; |\psi^{(2)}\rangle = \left(\frac{1+i}{2},\frac{1}{\sqrt{2}} \right)\; , \\
|\psi^{(3)}\rangle & = \left(\frac{1+i}{(\sqrt{3}-1)\sqrt{3+\sqrt{3}}},\frac{1}{\sqrt{3+\sqrt{3}}} \right) \; .
\end{align}
and $\vec{s}\,'$ can be readily obtained
\begin{align}
\vec{s}\,' & = \vec{s}\,'^{(1)} \vee \vec{s}\,'^{(2)} \vee \vec{s}\,'^{(3)} \nonumber \\
& = (1, \frac{\sqrt{2}}{2}, \frac{1+\sqrt{3}-\sqrt{2}}{2},\frac{1-\sqrt{3}+\sqrt{2}}{2}, \frac{2-\sqrt{2}}{2},0 )^{\mathrm{T}} \; .
\end{align}
Here $\vec{s}\,'$ is the unique least upper bound of the uncertainty relation $\vec{p}_x\oplus \vec{p}_z \oplus p_z\prec \vec{s}\,'$ for all the quantum states.

\subsection{Two observables in pure qutrit system}

We take the 3-dimensional observables $X$ and $Y$ with the orthonormal bases of \cite{Example-Coles-3}
\begin{align}
(|x_1\rangle, |x_2\rangle, |x_3\rangle) &= \begin{pmatrix}
1 & 0 & 0 \\
0 & 1 & 0 \\
0 & 0 & 1
\end{pmatrix} \;, \\
 (|y_1\rangle, |y_2\rangle, |y_3\rangle) &=
\begin{pmatrix}
\frac{1}{\sqrt{3}} & \frac{1}{\sqrt{3}} & \frac{1}{\sqrt{3}} \\
\frac{1}{\sqrt{2}} & 0 & -\frac{1}{\sqrt{2}} \\
\frac{1}{\sqrt{6}} & -\sqrt{\frac{2}{3}} & \frac{1}{\sqrt{6}}
\end{pmatrix} \; , \label{S-tri-Y}
\end{align}
as an example. For pure states, the largest possible one component in $\vec{\chi} = \vec{p}_x \oplus \vec{p}_y$ is $\max\{\tau_1\} = 1$. This is obtained by any one of quantum states in $\{|x_1\rangle, |x_2\rangle, |x_3\rangle, |y_1\rangle, |y_2\rangle, |y_3\rangle\}$. Taking $|x_1\rangle$, we have
\begin{align}
\vec{s}\,''^{(1)} = (1,\frac{1}{3}, \frac{1}{3},\frac{1}{3},0,0)^{\mathrm{T}} \; .
\end{align}
One may also take, for example $|x_3\rangle$, and we would get $\vec{s}\,''^{(1)} = \displaystyle (1,\frac{2}{3}, \frac{1}{6},\frac{1}{6},0,0)^{\mathrm{T}}$ which will give the same $\vec{s}\,''$ at last.

The largest sum of any two components of $\vec{\chi}$ correspond to finding the largest eigenvalue in $|x_i\rangle\langle x_i| + |y_j\rangle \langle y_j|, i,j \in \{1,2,3\}$. It is quite clearly from equation (\ref{S-tri-Y}) that $|x_3\rangle \langle x_3| + |y_2\rangle\langle y_2|$ has the largest eigenvalue $(3+\sqrt{6})/3$, and therefore $\max\{\tau_2\} = (3+\sqrt{6})/3$. The quantum state  giving the largest sum of any two components is just the eigenvector correspond to this eigenvalue
\begin{align}
|\psi^{(2)} \rangle = (- \sqrt{\frac{1}{2} - \frac{1}{\sqrt{6}}},0,\sqrt{\frac{1}{2} + \frac{1}{\sqrt{6}}} ) \; .
\end{align}
Taking $|\psi^{(2)} \rangle$ into $\vec{\chi}$, we have
\begin{align}
\vec{s}\,''^{(2)} = (\frac{1}{2}+\frac{1}{\sqrt{6}},\frac{1}{2}+\frac{1}{\sqrt{6}},\frac{1}{2}-\frac{1}{\sqrt{6}},\frac{1}{12} \left(3-\sqrt{6}\right),\frac{1}{12} \left(3-\sqrt{6}\right),0)^{\mathrm{T}} \; .
\end{align}

The largest sum of any three components of $\vec{\chi}$ corresponds the largest eigenvalue in $|x_i\rangle \langle x_i| + |x_j\rangle \langle x_j| + |y_k\rangle \langle y_k|$ or $|y_i\rangle \langle y_i| + |y_j\rangle \langle y_j| + |x_k\rangle \langle x_k|$ with $i\neq j$. Simple evaluation shows that $|x_1\rangle \langle x_1| + |x_3\rangle \langle x_3| + |y_2\rangle \langle y_2|$ has the largest eigenvalue $2$ with the eigenvector
\begin{align}
|\psi^{(3)} \rangle = \left(- \frac{1}{\sqrt{3}},0,\frac{\sqrt{2}}{\sqrt{3}} \right) \; .
\end{align}
Hence $\max\{\tau_3\} = 2$. Taking $|\psi^{(3)} \rangle$, we have
\begin{align}
\vec{s}\,''^{(3)} = (1, \frac{2}{3}, \frac{1}{3}, 0,0,0)^{\mathrm{T}} \; .
\end{align}

No further calculation is needed, because the largest sum of any three components already reaches the value $2$ where $p_{x_1}+ p_{x_2}+ p_{x_3}+ p_{y_1}+ p_{y_2}+ p_{y_3}=2$. We have
\begin{align}
\vec{s}\,'' & =  \vec{s}\,''^{(1)} \vee \vec{s}\,''^{(2)} \vee \vec{s}\,''^{(3)} \nonumber \\
&= (1,\frac{\sqrt{6}}{3}, 1-\frac{\sqrt{6}}{3},0,0,0)^{\mathrm{T}} \; ,
\end{align}
and we can readily get the optimal bound for the universal uncertainty relation
\begin{equation}
\vec{p}_x \oplus \vec{p}_y \prec \vec{s}\,''\; .
\end{equation}
Here $\vec{s}\,''$ is the unique least upper bound of the uncertainty relation $\vec{p}_x\oplus \vec{p}_z \prec \vec{s}\,''$ over all the quantum states.

\end{document}